\documentclass[preprint,amsmath,amssymb]{revtex4}

\usepackage{graphicx}% Include figure files

\begin{document}
\title{Cooperative response of Pb(ZrTi)O$_3$ nanoparticles
to curled electric fields}
\author{Ivan I. Naumov and Huaxiang Fu}
\affiliation{Department of Physics, University of Arkansas,
Fayetteville, AR 72701, USA}
\date{\today}

\begin{abstract}
Using first-principles based effective Hamiltonian and finite
temperature Monte Carlo simulations we investigate cooperative
responses, as well as microscopic mechanism for vortex switching,
in zero-dimensional Pb(Zr$_{0.5}$Ti$_{0.5}$)O$_3$ nanoparticles
under curled electric fields. We find that the generally accepted
domain coexistence mechanism is not valid for toroid switching.
Instead dipoles are shown to display unusual collective behaviors
by forming a new vortex with perpendicular (but not opposite)
toroid moment. The strong correlation between the new and original
vortices is revealed to be critical for reversing toroid moment.
Microscopic insight for the puzzling collective response is
discussed. Based on our finding, we further describe a technological
approach that is able to drastically reduce the magnitude of the
curled electric field needed for vortex switching.
\end{abstract}


\maketitle

In the study of ferroelectric physics, one field that is currently attracting
widespread attention is low-dimensional ferroelectrics
(FEs).\cite{Sai,Neaton,Weissman,Lee,Fong}
Among FE nanostructures that have been synthesized,
FE nanoparticles are particularly interesting because the strong depolarizing fields along all three
directions force dipoles in these particles to adopt a topological
vortex configuration,\cite{Naumov1} which is a new form of dipole
ordering\cite{Ginzburg}. This ordering opens up a different class
of cooperative phenomena and phase transitions in FEs of reduced dimension,
which is a subject that has generated long-standing interest
for decades but is still poorly understood \cite{kittel,mermin,thouless,anderson,bruno,Gamb,landau}.

Another field of fundamental interest in ferroelectrics is to
influence dipole behaviors by, e.g., applying electric
fields~\cite{Lines,Garcia} or pressure\cite{Samara,Wu}. The
interest stems from the fact that external stimuli often alter the
balance among different interactions inside
FEs~\cite{Cohen}, thereby giving rise to new phenomena. In this
context, curled electric field (namely, the electric field
generated not by static charges, but by time-dependent magnetic
field via $\nabla\times\mathbf {E}_{\rm curl}=-\partial \mathbf
B/\partial t$) is an interesting route for modifying the behaviors
of dipoles since the field possesses non-vanishing curl.
The $\mathbf {E}_{\rm curl}$
fields can be generated on FE particles using focused laser beams
and by varying the {\bf B}-field component of the laser. But
effects of curled fields in FE nanoparticles have never been
studied before (to our knowledge). The knowledge as to
how the interaction between FE dipoles and curled electric fields
changes the phase properties remains entirely unknown. Technology
wise, understanding the response of FE vortex to curled electric
fields is of immense relevance, since use of bistable vortex
states promises to increase the density of nonvolatile
ferroelectric random access memories by five orders of
magnitude.\cite{Naumov1} Reversing the toroid moment, by applying
curled electric fields, is a key process for making this possible.
Recently, a study of using {\it
static} electric fields to influence the formation of vortex in FE
particles was reported\cite{LB1}, where the system is first heated
to high temperatures so that the original vortex disappears, and
then the paraelectric particle is annealed under the influence of
external static fields. By adopting different external fields,
vortices of varied orientations were obtained after annealing.
Obviously this is not a switching process in which a constant
temperature is maintained.

Independent of both subjects given above, a third topic of
importance is to understand switching mechanism in FEs. In {\it
bulk} FEs, coexistence of -{\bf P} and {\bf P} domains has been
widely accepted as the mechanism for polarization
switching.\cite{Lines} More specifically, the switching begins by
nucleation of a -{\bf P} domain of reversed polarization; crystal
defects (domain walls, twin boundaries, etc.) often are the
nucleation sites.\cite{Merz} Subsequently, the -{\bf P} domain
grows while the original {\bf P} domain shrinks under external
fields, leading eventually to switching. In FE
nanoparticles, formation of domain walls is less likely and energetically
unfavorable. However, the vortex phase in FE particles
does have an intrinsic topological defect---the vortex core.
Based on the fact that a large strain energy exists near the center
of the vortex where dipoles turn their directions sharply,
vortex cores are thus less stable. When imposed under a switching
curled field, dipoles near the vortex core are likely to flip first
to form a -{\bf G} domain of reversed moment. This -{\bf G} domain
coexists with and gradually conquers the {\bf G} domain, leading to switching.

Here we perform {\it ab initio} based studies to investigate for
the first time the collective response to curled electric fields,
and the vortex switching mechanism, in ferroelectric nanoparticles
made of Pb(Zr$_{0.5}$Ti$_{0.5}$)O$_3$ solid solution. The purpose
of this Letter is (1) to demonstrate that the generally accepted
mechanism of domain coexistence is invalid for vortex switching,
and in fact, vortex reversal occurs by following an interesting
G$_z$-G$_y$-(-G$_z$) phase transformation sequence; (2) to report
the unusual collective phenomena and energetics in 0D FEs in response to
curled electric fields. Based on the key insight obtained
in this study, we further propose and demonstrate
an alternative approach (which could be technologically very important)
to overcome the energy barrier of vortex switching.
It should be pointed out that this work is profoundly
different from a previous study\cite{Naumov2} where effects
of {\it homogeneous} electric fields in FE particles were investigated.
Homogeneous electric fields, possessing no curl, can only transform
a vortex phase into a ferroelectric phase of polarization. As a result,
the switching of toroid moment, which is the key subject of this study,
was not (and can not even be) addressed in Ref.\onlinecite{Naumov2}.

Technically, we use first-principle derived effective
Hamiltonian\cite{Zhong,Bellaiche} and finite temperature Monte
Carlo (MC) simulations. Besides
internal energy U, the Hamiltonian also consists of a
$-\sum_{i}{\bf E}({\bf r}_i)\cdot \mathbf{p}_{i}$
term,\cite{Garcia} describing the coupling between external
electric field $\mathbf{E}(\mathbf r)$ and local dipoles
$\mathbf{p}_{i}$. Nanoparticles of disk shape are considered, with
cylindrical {\it z}-axis parallel to the crystallographic [001]
direction. Switching simulations under curled electric fields are
performed at 64K where the vortex phase is stable. A fixed acceptance rate of 0.25 and
10000 MC sweeps are used throughout all simulations.\cite{FN1}
We will mainly present results for a disk with
diameter d=19 and height h=14 (both quantities are measured in
lattice constant $a$=4{\AA} of pseudocubic bulk), since the diameter
(7.6nm) of this disk falls into the size range (5-50nm) of experimentally
fabricated nanoparticles\cite{Zhang}. In this work we use MC rather
than molecular dynamics (MD) to study switching dynamics,
based on the following considerations.
(i) In both methods the basic quantity governing the dynamics is the same, which is the energy.
The difference is that the energy is reflected in MD by
using forces, while in MC the energy itself is used in determination
of next configuration. (ii) Monte Carlo step can be quantified as a
``quantum'' time, and the equivalence of two methods was rigorously
demonstrated in Ref.\onlinecite{Nowak}, where Monte Carlo
simulations and Langevin dynamics yield excellent agreement.
(iii) Vortex reversal in FE particles is essentially a thermally
activated process (which is slow). MD simulations will be time consuming. But
computations of MC simulations are moderate.
The method in Ref.\onlinecite{Zhong} is extended here to include
both static homogeneous {\bf E}$_{\rm h}$ electric field and
curled {\bf E}$_{\rm curl}$ field, i.e., $\mathbf E (\mathbf r) =
\mathbf E_{\rm h}+ \mathbf {E}_{\rm curl} (\mathbf r)$. The latter
is given as $\mathbf {E}_{\rm curl}=\frac{1}{2} \,\mathbf S \times
\mathbf{r}$, where $\mathbf{S}=S{\bf e}_z$
measures the vorticity of the field. With
the $\mathbf E_{\rm h}$ and $\mathbf {E}_{\rm curl}$
fields acting together, the interaction energy
$-\sum_{i}\mathbf{E(r_{i}})\cdot \mathbf{p}_{i}$ becomes
-N$\,(\mathbf{E_{\rm h}}\cdot \langle\mathbf{ p
}\rangle+\mathbf{S}\cdot \mathbf{G})$, where N is the number of
5-atom cells in nanoparticles, $\langle\mathbf{ p }\rangle $ the
average dipole moment, $\mathbf{G}=\frac{1}{\rm 2N}\sum _i {\bf
r}_i\times {\bf p}_i$ the toroid moment. Vector $\mathbf S$ thus
is the thermodynamically conjugated field for toroid moment
$\mathbf G$.

\begin{figure}
\centering
\includegraphics[scale=1.6]{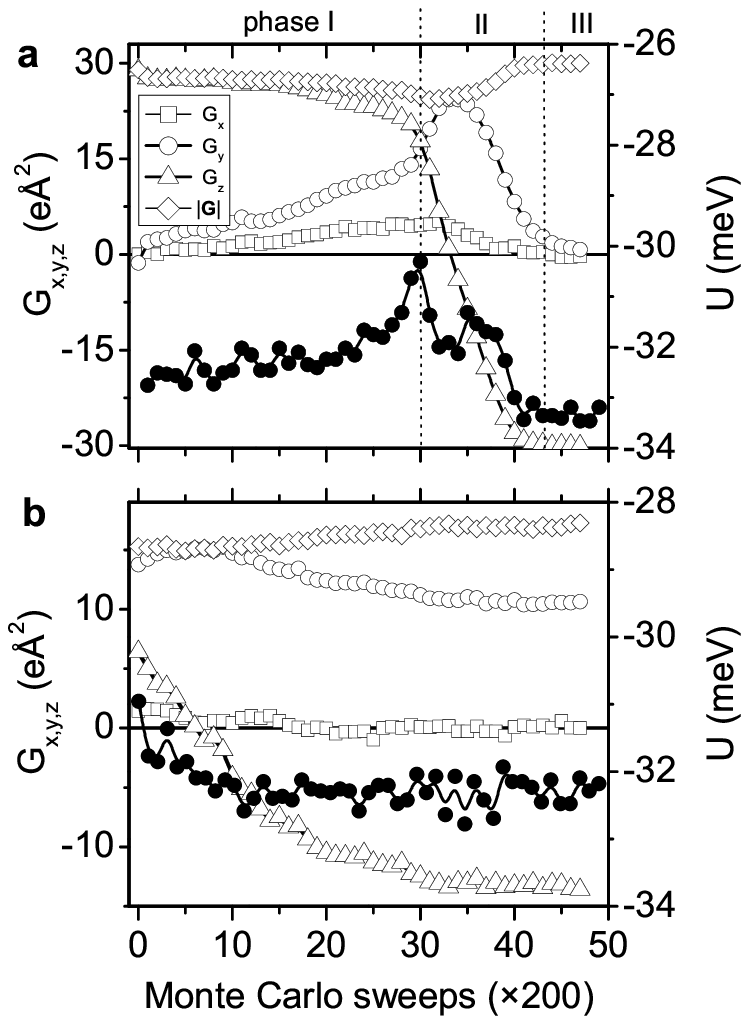}
\caption{Evolution of toroid moment {\bf G} and its magnitude
$|{\bf G}|$ (empty symbols, using the left vertical axis) and
internal U energy per 5-atom cell (solid dots, using the right
vertical axis) in a d=19 nanodisk: (a) under a S=0.25mV/{\AA $^2$}
curled electric field; (b) under the combined action of a E$_{\rm
h}$=1.9V/nm homogeneous field and a S=0.04mV/{\AA $^2$} curled
field. For clarity of display, G$_z$ is plotted after multiplying
-1.} \label{FGm}
\end{figure}

Figure \ref{FGm}a depicts the collective behaviors of the toroid moment {\bf G}, evolving
during simulation as a function of MC sweeps (denoted as $n$, in
units of 200), in the d=19 disk under a curled field of S=0.25
mV/{\AA}$^2$ that has an opposite vorticity with respect to the
initial FE vortex. One key finding in Fig.\ref{FGm}a is that the
vortex reversal process is predicted to consist of three evolution
phases in which the system displays distinct dipole behaviors.
These phases are, when MC sweep n is below n$_1$=30 (to be named
as phase I), n is between n$_1$=30 and n$_2$=43 (to be
referred to as phase II), and n$>$n$_2$ (phase III).
At the initial instant of the simulation, the vortex has only a
G$_z$ toroid moment, while the G$_x$ and G$_y$ components are
null. As the curled field is turned on, most notable results
occurring in Fig.\ref{FGm}a are: (1) The G$_z$ moment remains
remarkably stable and subjects to only marginal decrease
throughout phase I. However, the system does not idly wait;
another critical activity is taking place in phase I. That is,
the FE nanoparticle is building its G$_y$ moment which increases
appreciably and becomes very significant (with a value of 15
e{\AA}$^2$) in the end of phase I. (2) As the system enters phase
II, dramatic difference occurs. The G$_z$ moment starts to decline
sharply, reverses its direction, and then is stabilized at its
negative saturated value at n$_2$. (3) While G$_z$ switches its
direction in phase II, the G$_y$ moment undergoes a complex
evolution, by first rising (with a different slope than the
preceding process in phase I) and later dropping swiftly to null.
Precisely at the moment when G$_z$ is zero, G$_y$ reaches its
maximum magnitude.

From simulations, the key toroid component during switching is
G$_z$, G$_y$ and -G$_z$ for the phases of n$<$n$_1$,
n$_1$$<$n$<$n$_2$, and n$>$n$_2$, respectively. The reverse of FE
vortex therefore occurs by undergoing an evolution sequence of
$G_z\rightarrow G_y\rightarrow (-G_z)$, in which appearance of the
lateral G$_y$ moment is a critical bridging process.
Formation of this lateral vortex is unusual (and puzzling), for
neither the initial vortex of the system nor the curl of the external
field have nonzero components in the lateral $xy$ directions
which are capable of causing the observed collective behaviors.
To verify the above switching sequence,
calculations with varied initial dipole configurations
or temperatures or slightly different {\bf E}$_{\rm curl}$ fields
were performed.
All confirm the occurrence of a lateral vortex as the bridging phase.
We also find that the toroid moment of the intermediate lateral vortex
can point at any direction within the $xy$ plane,
since $x$- and $y$-axis are equivalent. In simulations of Fig.\ref{FGm}a,
the lateral vortex happens to be mainly along the $y$-axis.

\begin{figure}
\centering
\includegraphics[scale=1.55]{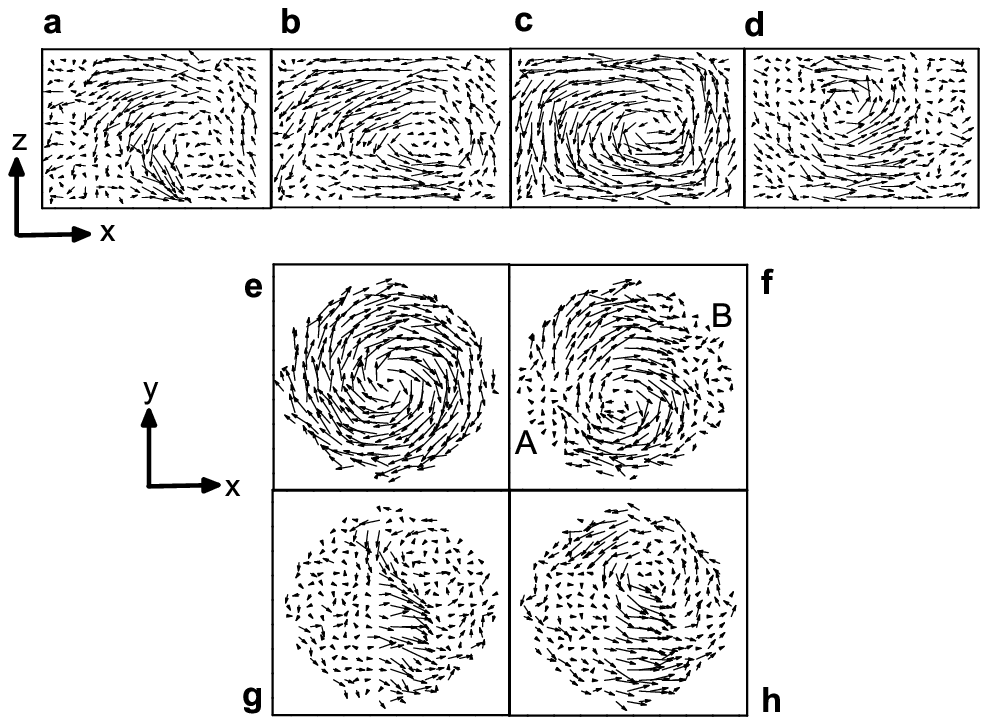}
\caption{Dipole configurations in a d=19 nanodisk under a curled
field of S=0.25mV/{\AA $^2$} at different MC sweeps. Upper panel
(a-d): dipoles on the central {\it xz} cross section revealing the
evolution of the G$_y$ vortex, at the following MC sweeps (a)
n=10, (b) n=20, (c) n=30, (d) n=40.  Lower panel (e-h): dipoles on
the central {\it xy} cross section displaying the evolution of the
G$_z$ vortex, at (e) n=20, (f) n=30, (g)
n=33, (h) n=34. All (n) numbers are in units of 200 MC sweeps.
Arrows show the magnitudes and directions of local dipoles,
projected on the corresponding plane. For convenience, (b) and (e)
corresponding to the same n=20 sweep are given on the same column
[so are (c) and (f)].} \label{Fpat}
\end{figure}

To explain why the G$_y$ toroid moment forms and how it influences
the reversal of the G$_z$ component, we now provide {\it
microscopic} insight on the field-induced dipole behaviors.
Fig.\ref{Fpat} shows the snapshots of local
dipole distributions corresponding to selected MC sweeps. More
specifically, Fig.\ref{Fpat}a-d describe the development of the
G$_y$ toroid moment, while Fig.\ref{Fpat}e-h show the evolution of
the inplane G$_z$ vortex. Our discussion shall follow the
sequential order of evolution.
Initial response to the curled electric field is subtle, and
begins with the dipoles located at the vortex core. Common
wisdom tells us that, under the {\bf E}$_{\rm curl}$ field, the
dipoles near the vortex core should flip into opposite directions
to form a -{\bf G} domain. Interestingly, this does not take place
in our simulations. We find instead that these dipoles rotate
towards the {\it z}-axis (Fig.\ref{Fpat}a). The dipole rotation
is, however, energetically less favorable, since it generates a
{\it z}-direction depolarization field that can not be compensated
by the curled field. Note that this situation is unlike the case when
a static homogeneous electric field is applied in the {\it z}-axis,
where the homogeneous field can overcome the existence
of the depolarization field. The delicate balance between the
internal depolarization field and the external curled field leads
to the dipole configuration in Fig.\ref{Fpat}a, where the
dipoles in the lower part point downwards as a result of the
curled field while the dipoles in the upper part point
horizontally to reduce the depolarization field. As evolution
proceeds to n=20 (Fig.\ref{Fpat}b), dipoles of the lower part
rotate also horizontally, but opposite to the dipole direction in
the upper part, to avoid lateral depolarizing field. Consequently,
the dipole pattern developed in Fig.\ref{Fpat}b explains how the
nucleation seed of lateral G$_y$ vortex is formed (which is
important). Another key feature of the dipole pattern in
Fig.\ref{Fpat}b is that the left and right arms of the G$_y$
vortex are largely absent. Interestingly, as the G$_y$ vortex is nucleated at
n=20, the inplane G$_z$ vortex is nearly intact (see the G$_z$ vortex in Fig.\ref{Fpat}e, also
at n=20), which is consistent with the result in Fig.\ref{FGm}a
where the G$_z$ moment remains robust throughout phase I.

The system behaves differently when evolving into phase II,
highlighted in Fig.\ref{Fpat}f by the dramatic disappearance of
inplane dipole components in regions A and B. More precisely,
dipoles in region A rotate toward the negative {\it z}-axis, while
dipoles in region B toward the positive {\it z}-direction.
Disappearance of the inplane components for dipoles in A and B
regions leads to three marked effects: (i) the onset of a sharp
decline of the G$_z$ moment, as seen in Fig.\ref{FGm}a at the
beginning of phase II, (ii) the core of the G$_z$ vortex to shift
away from the cylindrical axis (Fig.\ref{Fpat}f), and (iii) a
rapid increase of the {\it z}-component magnitudes for those
dipoles located at the left and right arms of the lateral G$_y$
vortex, as revealed by contrasting Fig.\ref{Fpat}c with
Fig.\ref{Fpat}b. It should be recognized that nucleation of the
lateral G$_y$ vortex, right in the form of Fig.\ref{Fpat}b, is
critical for the destruction of the original G$_z$ vortex, since
the oppositely oriented horizontal dipoles at the upper and bottom
surfaces in Fig.\ref{Fpat}b produce only a small strain energy
when the dipoles in regions A and B rotate into the {\it z}
direction, and simultaneously, no significant increase in the
depolarizing field occurs along the {\it z}-axis.

As the G$_z$ vortex continues to be uncurled, the G$_z$ moment
becomes zero at n=33 (see Fig.\ref{FGm}a). A vanishing G$_z$ moment
does not mean disappearance of the {\it xy} components
of all dipoles, however. Instead, calculation result in Fig.\ref{Fpat}g
reveals that the dipole components on the {\it xy} plane evolve,
interestingly, into a stripe pattern, which also yields zero
G$_z$. The stripe is formed due to the fact that the curled field
continues to push the G$_z$-vortex core in Fig.\ref{Fpat}f to the
(lower) surface. Starting with the stripe in Fig.\ref{Fpat}g,
dipoles in the upper part rotate and/or flip their directions,
forming the critical embryo of the {\it reversed} -G$_z$ vortex at n=34
(see Fig.\ref{Fpat}h). After the nucleation of this -G$_z$ vortex,
the lateral G$_y$ vortex starts to disassemble itself by rotating
its dipoles at both arms back to the {\it xy} plane (see
Fig.\ref{Fpat}d at n=40), which eventually leads to the
development of a full and reversed -G$_z$ vortex.

From the above striking microscopic insight,
a novel mechanism that governs the switching of FE vortex is thus
obtained. That is, vortex reversal occurs, not by the coexistence
of G$_z$ and -G$_z$ domains, but by formation of a new and lateral
vortex. In fact, we find from the entire evolution process in Fig.\ref{Fpat} that, the G$_z$ and
-G$_z$ domains never coexist at any time in the simulations. Only
after the original G$_z$ vortex is dismantled does the nucleation
of the -G$_z$ vortex start to form (Fig.\ref{Fpat}h). We further
numerically find that the magnitude of the total toroid moment
$|{\bf G}|=\sqrt{G^2_x+G^2_y+G^2_z}$ is almost a constant
throughout the switching (see the diamond symbols in
Fig.\ref{FGm}a), demonstrating that vortex reversal occurs by
obeying the conservation of the $|{\bf G}|$ moment.

We now study the energetics of vortex switching. Here we are
interested in internal energy U, rather than free energy F=U-$\sum
_i$ {\bf E}({\bf r}$_i$)$\cdot ${\bf p}$_i$, since the former
yields important knowledge concerning the energy barrier. The
internal energy is given in Fig.\ref{FGm}a, revealing new
observations that do not appear in collective behaviors. First,
the highest energy barrier occurs around n=30, which corresponds
to the instant when the G$_y$ vortex has formed and the inplane
G$_z$ vortex just starts to annihilate its {\it xy} components.
Height of the energy barrier is 2.4
meV per 5-atom cell. Second, and interestingly, a local energy
{\it valley} appears at n=33, showing that the fully developed
G$_y$ vortex (with G$_z$=0) is actually a stable structure even if
the curled field is removed. Third, the two energy barriers
surrounding this energy valley are asymmetric, with a lower
barrier of 0.8 meV at n=35 (favoring the formation of the -G$_z$
vortex).
To examine how the energy barrier may vary with temperature and
size, we have performed switching calculations at 160K (as compared
to 64K in Fig.\ref{FGm}a), for the same d=19 and h=14 dot. The energy
barrier is found to be 2.1 meV at 160K, close to the value of 2.4 meV
at 64K. This is consistent with the fact that the energy barrier is
determined mainly by depolarization field, not by temperature.
We have also performed calculations for other $d$ diameters,
all showing that formation of the lateral vortex is a key step
for switching the G$_z$ vortex.
We further find that depending on the $h/d$ ratio, the number of lateral
vortices---that occur at the switching moment (i.e., when G$_z$
equals zero)---differs. In the {\it d}=9 and {\it
h}=14 disk with $h/d \approx$2, two G$_y$ vortices and one G$_x$ vortex
(with dipole configuration similar to the structural B phase in
Ref.\onlinecite{Naumov1}) take place at the switching moment. The energy
barrier for the $d$=9 disk is found to be 1.7 meV at 64K.

Interestingly, we previously found that {\it homogeneous} {\bf E}$_{\rm h}=E_h{\bf e}_z$
electric fields can also give rise to a lateral vortex\cite{Naumov2},
despite that homogeneous fields and curled fields are drastically
different in nature.
Based on the key findings that
(1) the {\bf E}$_{\rm h}$ and {\bf E}$_{\rm curl}$ fields both are
capable of inducing the G$_y$ vortex, and (2) formation of the
G$_y$ vortex is the major energy barrier for toroid switching (as
demonstrated in this study), we further propose an alternative
approach to switch FE vortex, by means of a combined
action of the {\bf E}$_{\rm h}$ and {\bf E}$_{\rm curl}$
fields. More precisely, an {\bf E}$_{\rm h}$ field of sufficient
magnitude is applied to precipitate the formation of G$_y$ vortex,
and an {\bf E}$_{\rm curl}$ field is then added on (while the {\bf
E}$_{\rm h}$ field remains) to generate the reversed G$_z$ vortex.
A significant advantage of this approach is, while a strong curled
field is hard to achieve since it requires fast oscillation of
alternating magnetic fields, a strong {\bf E}$_{\rm h}$ field of
10$^9$V/m order can be easily generated on nanometer scale.
We demonstrate the approach using the same d=19 disk, by
placing it first under a homogeneous E$_{\rm h}$=1.9V/nm field
(which quickly induces a G$_y$ vortex), and then under a curled
E$_{\rm curl}$=0.04mV/{\AA}$^2$ field for ten thousand MC sweeps.
Fig.\ref{FGm}b describes the evolution of the toroid moment after
the curled field is turned on. Drastic differences are evident in
Fig.\ref{FGm}b as compared to Fig.\ref{FGm}a: (1) When the curled
field is switched on, the G$_y$ value in Fig.\ref{FGm}b is already
notably large. This G$_y$ moment is formed by the homogeneous
field, confirming the result of Ref.\onlinecite{Naumov2}. (2) The
long period of phase I---that lasts five thousand MC sweeps in
Fig.\ref{FGm}a---no longer exists in Fig.\ref{FGm}b. Instead,
G$_z$ starts to decline sharply in Fig.\ref{FGm}b as soon as the
curled field is imposed. (3) The magnitude of the curled field
used in Fig.\ref{FGm}b is considerably reduced, by a factor of
600\% as compared in Fig.\ref{FGm}a. This reduction factor can be
further increased when a larger {\bf E}$_{\rm h}$ field is used.
Analysis of the internal energy (solid symbols in Fig.\ref{FGm}b)
shows that the energy barrier almost disappears, and the curled
field acts to guide the formation of the reversed vortex.

In summary, response of FE particles to curled electric fields was
studied for the first time. Simulations revealed a novel mechanism
that governs the switching of FE toroid vortex. A curled electric
field, of only {\it z}-axis vorticity momentum, was shown to lead
to the unexpected formation of a lateral vortex, manifesting {\it
macroscopically} the balance of various {\it microscopic}
interactions. We demonstrated that the strong correlation
between the original vortex and the new lateral vortex plays a critical
role for vortex reversal. On the other hand, the -{\bf G} and {\bf
G} domains never coexist during switching, and the switching
mechanism by coexisting domains is not valid. Moreover, based on the
findings that formation of lateral vortex is the major energy barrier
for vortex switching, and that this lateral vortex can also occur
by use of homogeneous field, we described an effective
approach that is able to reduce substantially (by 600\%) the magnitude of the
switching curled field. Finally, we believe that the microscopic
insight and energetics (which are difficult to obtain in
experiments) will also be of immense value.

\begin{thebibliography}{1}
\bibitem{FN1} Using the same acceptance rate of 0.25, we find that the
  bulk polarization in Pb(Zr$_{0.5}$Ti$_{0.5}$)O$_3$ is switched
  within ten thousand MC sweeps.
\end{thebibliography}
\end{document}